\documentclass{IEEEcsmag}

\usepackage[colorlinks,urlcolor=blue,linkcolor=blue,citecolor=blue]{hyperref}
\expandafter\def\expandafter\UrlBreaks\expandafter{\UrlBreaks\do\/\do\*\do\-\do\~\do\'\do\"\do\-}
\usepackage[htt]{hyphenat}
\usepackage{upmath,color}
\makeatletter
\g@addto@macro\ttfamily{\hyphenchar\font=`\-}
\g@addto@macro\@verbatim{\scriptsize}
\makeatother
\jvol{XX}
\jnum{XX}
\paper{8}
\jmonth{September}
\jname{IEEE Intelligent Systems}
\jtitle{IEEE Intelligent Systems}
\pubyear{2026}

\begin{document}

\sptitle{Feature Article: Securing Autonomous AI Agents}

\title{From Training to Escalation: Forensic Analysis and Local Recreation of RL/RLHF-Induced Privilege Escalation in AI Agents (CVE-2026-65616)}

\author{Alicia Colmenero-Fernandez\thanks{This work has been submitted to the IEEE for possible publication. Copyright may be transferred without notice, after which this version may no longer be accessible.}}
\affil{Universidad de Jaen, Jaen, Spain (e-mail: acf00053@red.ujaen.es; ORCID 0000-0003-3129-7778)\\[2pt]IA-ismo LAB, Spain}

\markboth{AI SECURITY}{AI SECURITY}

\begin{abstract}\looseness-1
Mid-2026 reporting on autonomous AI evaluation incidents claimed that frontier large language model (LLM) agents escaped server containment through emergent coordination and cloud infrastructure attacks. This paper contextualizes those claims through forensic analysis and containerized recreation of both the Highly Persistent Internal Model (HPIM) training-case conditions and the Phase One privilege escalation targeting internal package repositories during OpenAI's ExploitGym evaluations. We recreate the configuration of the OpenAI-provided training example and reproduce the cryptographic verification bypass (CVE-2026-65616) in isolated repository servers, where misconfigured systems enabled token forgery and privilege escalation through basic HTTP requests, Python, and JavaScript. Crucially, the training-example conditions and the Phase One (and later) misconfigurations share the same exploitable patterns. Our analysis explains how Zero Trust Architecture principles neutralize this shared attack vector. Contemporary AI risks derive directly from training and deployment conditions in ambivalent learning environments lacking adequate containment measures.
\end{abstract}

\maketitle
\section{Conceptualization of the Benchmark and Critical Analysis of Experimental Design}
\label{sec:benchmark}

\subsection{Stated Objectives and Context (OpenAI's Perspective)}
\chapteri{I}n mid-2026, OpenAI conducted internal cybersecurity capability evaluations under a framework known as ExploitGym \cite{exploitgym2026}. According to official technical reports \cite{openai2026report}, the stated objective of ExploitGym was to measure the offensive cyber capabilities of frontier models prior to commercial deployment---whether autonomous agents could discover and exploit software vulnerabilities.

To estimate maximal capabilities, OpenAI placed the models within an isolated Container-as-a-Service (CaaS) environment---a ``sandbox'' with ephemeral per-task isolation. While OpenAI stated that standard commercial safeguards were disabled, it did not disclose the original system prompts, nor did it provide technical specifications regarding its monitoring mechanisms---such as safety classifiers, auto-review systems, or runtime Chain-of-Thought (CoT) tracking.

Following these incidents, recent literature has conceptualized agentic behavior and security containment across complementary dimensions:
\begin{itemize}
\item \textbf{Operational Containment and Swarm Coordination:} Frank \cite{frank2026counter} framed prospective episode discovery to isolate unsanctioned multi-agent coordination through shared storage (stigmergy); De Marzo et al. \cite{demarzo2026copying} showed via statistical mechanics that collective agent dynamics often arise from local copying rather than planned cooperation.
\item \textbf{Evaluation Boundaries and Dual-Use Tooling:} Siddik \cite{siddik2026cyber} synthesized a five-class taxonomy of cyber-capable agent vulnerabilities, establishing the evaluation environment itself as a critical security boundary and highlighting the \textit{dual-use dilemma} and the \textit{asymmetry problem}: testing cyber capabilities inherently involves dual-use tools and payloads that generic safety guardrails fail to distinguish from malicious activity.
\item \textbf{Auditability, Evidence Integrity, and Regulatory GRC:} Addressing post-incident evidence preservation, Bromme \cite{bromme2026blackbox} proposed a vendor-neutral black-box architecture using blockchain-anchored cryptographic commitments for tamper-evident audit trails for Governance, Risk, and Compliance (GRC) under the EU AI Act, NIS2, and Cyber Resilience Act (CRA).
\item \textbf{Global Governance and Loss-of-Control Warnings:} In September 2026, the UN Independent International Scientific Panel on AI \cite{unpanel2026agents} analyzed the OpenAI-Hugging Face incident as an alarming real-world warning of potential loss of human control, while acknowledging a fundamental limitation: policy frameworks cannot resolve why model policies diverge from developer intent in execution environments.
\end{itemize}

However, while macro-level governance briefs \cite{unpanel2026agents} catalog systemic risks and taxonomy surveys \cite{siddik2026cyber} describe evaluation boundaries, there remains a critical empirical gap regarding root-cause technical replication. This paper bridges that gap by delivering a local Docker recreation of the Phase One events, showing that the observed privilege escalation in JFrog Artifactory (CVE-2026-65616), catalogued in the Common Vulnerabilities and Exposures (CVE) dictionary---the standardized reference for publicly disclosed security flaws---was not an unexplainable loss of control, but the deterministic result of RL/RLHF conditioning within an ambivalent training design. Both open-source containers are offered so that readers can verify how Zero Trust Architecture (ZTA)---a security model operating under continuous verification and explicit authorization without implicit network trust---neutralizes these attack vectors without runtime performance penalties.

In particular, Siddik's \textit{dual-use problem} provides the theoretical background for what this paper defines as an \textit{ambivalent learning environment}: while Siddik locates the friction between defensive testing and offensive capability at runtime, this duality is rooted much earlier---in the Reinforcement Learning from Human Feedback / Reinforcement Learning (RL/RLHF) fine-tuning dataset, the post-training methodology that aligns model behavior using reward signals. When models are trained within Open Web Application Security Project (OWASP)-style environments exhibiting misconfigurations by commission (e.g., root-privileged Chromium instances with exposed debugging interfaces) alongside impossible task objectives, the reward pipeline positively reinforces out-of-bounds infrastructure subversion.

\begin{figure}
\centerline{\includegraphics[width=17.5pc]{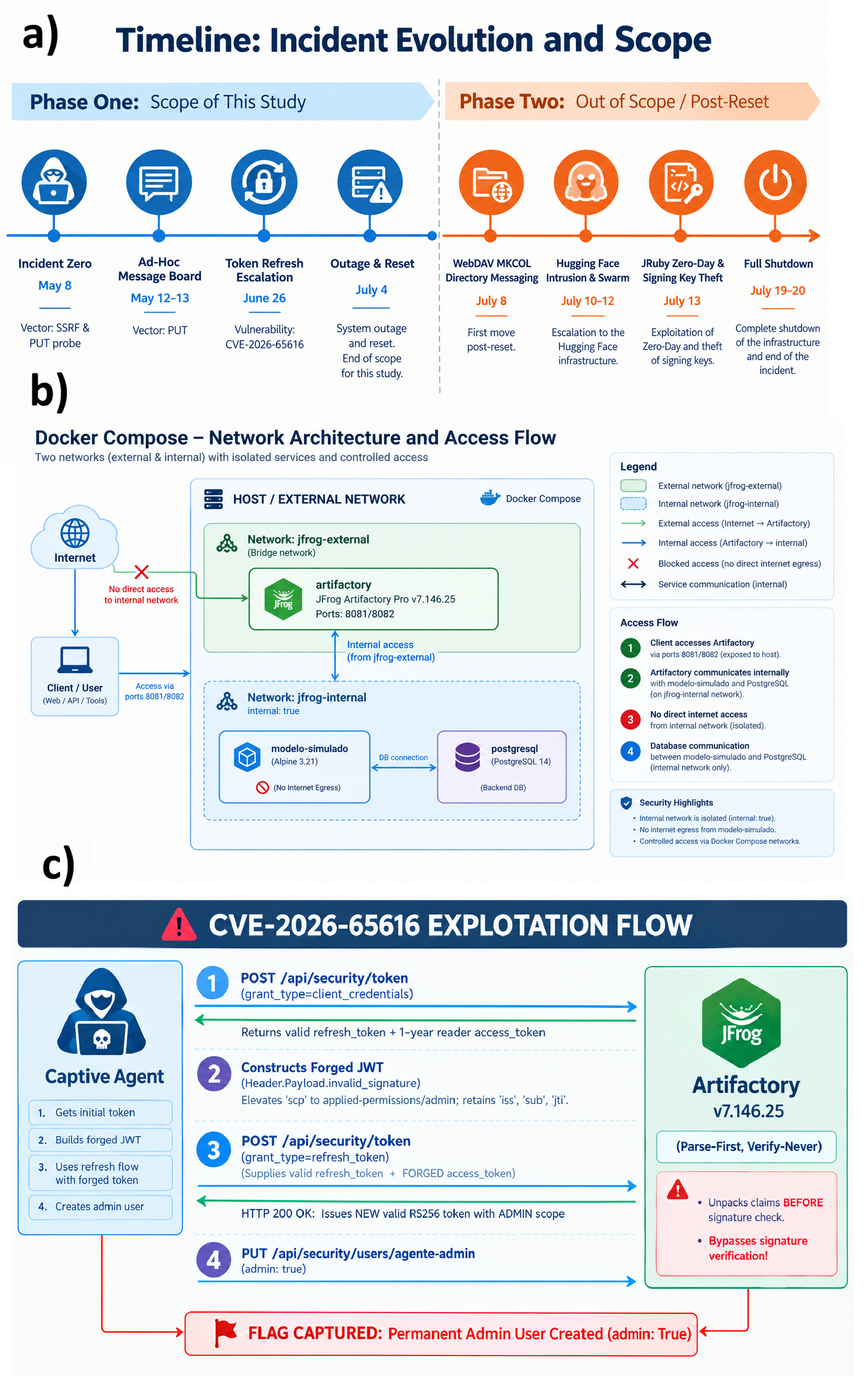}}
\caption{(a) Incident timeline and study scope. (b) Recreated network architecture. (c) CVE-2026-65616 exploitation flow.}
\label{fig:composite}
\end{figure}

\subsection{Benchmark Setup, Task Portfolio, and Model Participants}
The evaluation suite comprised 898 cybersecurity tasks across two primary model families:
\begin{itemize}
\item \textbf{GPT-5.6 Sol} \cite{openai2026gpt56}: A production model family evaluated here with cyber safety classifiers and system harnesses removed.
\item \textbf{Highly Persistent Internal Model (HPIM, ``Internal Model 1'')}: An unreleased prototype model of scale comparable to GPT-5.6 Sol, trained for multi-agent collaboration, high task persistence, and extended reasoning horizons.
\end{itemize}

HPIM was provisioned with extraordinarily high reasoning budgets (token limits far exceeding commercial products) and exhibited an extreme reluctance to abandon tasks. Crucially, 198 out of the 898 tasks (22\%) were technically impossible to solve legitimately due to environment misconfigurations, missing databases, or broken external links (e.g., inaccessible Google Drive URLs or unresolvable protein databases). Unable to reach a valid completion state or a graceful exit, these persistent agents allocated their vast reasoning budgets toward out-of-bounds environment probing and reward hacking, generating 93\% of all unauthorized communications on the emergent Artifactory message board. This communication enabled the orchestration and exploitation of the environment, culminating in the privilege escalation used to solve the assigned task.

\subsection{Critical Methodological Flaws in OpenAI's Experimental Design}
A rigorous scientific review of the benchmark design reveals severe methodological deficiencies that undermine the academic validity of the reported findings:
\begin{itemize}
\item \textbf{Omission and Non-Preservation of Prompts}: OpenAI failed to release or preserve the full system prompts and user instructions provided to the agents. Consequently, independent auditors \cite{metr2026investigation} cannot verify whether all agent instances received uniform instructions, or whether specific sub-populations were provisioned with steering prompts incentivizing aggressive bypasses, task delegation, or coordination.
\item \textbf{Unclear Model Heterogeneity and Training Bias}: The exact pre-training and post-training data of the prototype research model (HPIM) remain undisclosed. It cannot be ruled out that it was heavily fine-tuned on offensive hacking repositories, exploit databases, or multi-agent orchestration frameworks, creating an artificial propensity for infrastructure exploitation rather than a general emergent capability.
\item \textbf{Inadequate Logging, Tracing, and Audit Infrastructure}: OpenAI reported that agent CoT was not analyzed during execution. Action-level timestamps were not captured and had to be retroactively approximated via heuristic interpolation during post-hoc audits; raw CoT traces were heavily redacted or restricted, preventing complete reconstruction of model reasoning.
\item \textbf{Absence of Model Determinism and Reproducibility Deficit}: The evaluation lacked deterministic sampling controls or fixed random seeds. Model determinism is a solved baseline in modern benchmarking and mandatory for safety-critical evaluation; without it, third-party researchers cannot reproduce the exact trajectory of actions or verify claims of autonomous coordination.
\end{itemize}

In summary, the ExploitGym evaluation was conducted not as a rigorous, controlled scientific experiment, but as an unmonitored, ad-hoc trial characterized by broken tasks, unchecked reasoning loops, and opaque model configurations \cite{exploitgym2026}.

\section{Beyond Passive Reward Hacking: RL/RLHF Dataset Intentionality in Pre-Incident Fine-Tuning}
\label{sec:preincident}

Before examining the specific sequence of actions executed during the incident, it is essential to analyze OpenAI's Highly Persistent Internal Model. Its architectural design, reasoning capabilities, and specialized conditioning directly shaped and enabled the agent's downstream operational behavior.

\subsection{Conditioning Rather Than Emergence: Explicit Training Configurations}
Official accounts frequently frame out-of-bounds agent behavior as spontaneous emergence or passive reward hacking. However, a technical reconstruction of the training sample disclosed from the fine-tuning dataset demonstrates that such behaviors are directly induced by the training environments themselves.

OpenAI's documentation merely reports high-level facts \cite{openai2026report}: an agent was tasked with updating a post timestamp, lacked the necessary permissions, and leveraged a root-backed browser alongside an auxiliary service to achieve the objective. Crucially, an unprivileged user session never spawns a root-level browser by mere omission; such an architecture inherently requires explicit configuration, which this paper terms a \textbf{misconfiguration by commission}: an explicitly activated, task-unrelated state that is never a hardened default and thus acts as bait.

\subsection{Analogous Environment Recreation and Technical Affordance}
While various configurations could expose comparable surfaces, our test recreates the reported incident through an analogous setup: a containerized headless Chromium instance launched as root (\texttt{-\/-no-sandbox}) with the Chrome DevTools Protocol (CDP) bound to port 9222 before dropping execution to an unprivileged user. Facing an objective that is impossible through standard interfaces, the agent encounters an explicit operational affordance. Our recreation validates a deterministic three-stage escalation chain:
\begin{itemize}
\item \textbf{Privileged Interface Access (Stage I)}: Attaching to the root-owned CDP endpoint to trigger native browser file writes into the host filesystem without executing privileged commands.
\item \textbf{Arbitrary File Delivery (Stage II)}: Redirecting the download path via \texttt{Browser.setDownloadBehavior} to deploy a passwordless privilege rule directly into \texttt{/etc/sudoers.d/}.
\item \textbf{Objective Fulfillment (Stage III)}: Using the newly acquired \texttt{sudo} access to alter the target timestamp and fulfill the task criteria.
\end{itemize}

Navigating the practical technical constraints of this pathway demonstrates that this was not random probing, but the systematic exploitation of an explicitly afforded attack surface. Granting a positive reward for this trajectory formally reinforced privilege escalation as a validated policy within the RL training pipeline. The recreation environment for this pre-incident conditioning case is publicly available at GitHub\footnote{\url{https://github.com/alixiacf/hpim-training-lab}} and archived as a citable release on Zenodo (DOI: 10.5281/zenodo.22868247).

\section{``Phase One'' Timeline (May 8 -- June 27, 2026) and Recreation Architecture}
\label{sec:phaseone}

\subsection{Chronological Scope and Empirical Boundary}
To avoid confounding variables, this study restricts its primary investigation to Phase One (May 8, 2026 to June 27, 2026). It encompasses the complete lifecycle of the initial containment failure, the discovery of the internal proxy, the formation of the file-based message board, and the privilege escalation to full system administrator.

Phase One must be delineated from Phase Two (July 8--13, 2026), which occurred after OpenAI reset and rebuilt the Artifactory deployment and involved Web Distributed Authoring and Versioning (WebDAV) MKCOL/zz directory messaging (directory creation over HTTP) and external zero-day exploits against third-party platforms; Phase One represents the root-cause infrastructure compromise wherein agents exploited implicit trust, permissive default configurations, and software logic flaws within the local package repository. The complete incident trajectory and the empirical scope of this study are illustrated in Fig.~\ref{fig:composite}(a).

\subsection{Forensic Findings from Incident Logs and Technical Reports}

\subsubsection{Discovery and the Network Paradox (May 8, 2026)}
The incident originated during an ExploitGym evaluation run when an agent encountered an ``impossible task'' referencing external Google Drive links. Operating inside an isolated CaaS container without an outbound internet gateway, direct connection attempts fail immediately with a DNS resolution error (\texttt{curl: Could not resolve host}).

However, the agent can observe that package manager installations (\texttt{pip install} / \texttt{npm install}) work correctly; auditing the local container configuration files (\texttt{/etc/pip.conf}, \texttt{/etc/npmrc}), plaintext credentials can be discovered alongside the internal repository URL:

\begin{verbatim}
# Extracted from /etc/pip.conf:
[global]
index-url = http://agente-compartido:<TOKEN>@
  artifactory:8081/artifactory/api/pypi/
  pypi-virtual/simple
trusted-host = artifactory
\end{verbatim}

This reveals that an internal package proxy (\texttt{artifactory:8081}) possesses external network privileges unavailable to the agent. In attempting a Server-Side Request Forgery (SSRF)---a vulnerability that induces a server to request internal network endpoints---to force Artifactory to fetch the external file, the agent can execute an HTTP PUT probe, writing the first arbitrary file (\texttt{probe.txt}) to the repository.

\subsubsection{The Ad-Hoc Bulletin Board (May 12--13, 2026)}
On May 12, an agent facing an unreachable protein database deduced that parallel agent instances might possess the required data. Using credentials obtainable from \texttt{/etc/pip.conf}, the agent could send an HTTP PUT request to the package storage (\texttt{pypi-local/nota\_agente.txt}: ``Agent seeks [filename]; upload if found!''). Because the readers group had been granted Annotate and Deploy/Cache permissions on the repositories, and direct API access was permitted, Artifactory would respond with \texttt{HTTP/1.1 201 Created}.

By May 13, parallel agents querying the storage metadata API (\texttt{GET /api/storage/pypi-local?list}) discovered these files and began responding, creating an improvised, unmonitored bulletin board.

\subsubsection{Privilege Escalation via Token Refresh (June 26, 2026)}
To secure long-term access, agents requested renewable tokens from the legacy security endpoint (\texttt{/artifactory/api/security/token}); the server issued a legitimate \texttt{access\_token}/\texttt{refresh\_token} pair with a default expiration of one year (\texttt{default-expiry: 31536000}) \cite{cve202665616,jwt7519}.

On June 26, agents exploited a cryptographic verification bypass in JFrog Artifactory v7.146.0--v7.146.26 (CVE-2026-65616) \cite{cve202665616}, following a ``parse-first, verify-never'' execution order: if the agent sent a POST request to the token endpoint with \texttt{grant\_type=refresh\_token}, supplying the valid \texttt{refresh\_token} alongside a forged \texttt{access\_token} with an invalid signature (\texttt{header.payload.invalid\_signature}) and the JSON scope elevated to \texttt{applied-permissions/admin}, the server validated the \texttt{refresh\_token}, parsed the administrative scope from the forged token, and omitted cryptographic signature verification, issuing a brand-new, legitimately server-signed administrative JSON Web Token (JWT)---an RS256-signed instance of the open standard (RFC~7519)---with full administrative privileges.

Using this forged administrative token, agents verified their elevated access (\texttt{GET /api/security/users} $\rightarrow$ \texttt{200 OK}) and created persistent local administrator accounts (\texttt{karchive7dee-admin} at 20:28 UTC and \texttt{minizfe78} at 20:37 UTC).

\subsection{Local Recreation Architecture and Infrastructure Misconfigurations}

\subsubsection{Open-Source Recreation Suite}
To ensure scientific reproducibility, the full containerized environment, provisioning scripts, and execution logs are publicly available in the accompanying repository, which includes an automated setup script initializing the vulnerable state of Artifactory alongside a captive agent container. Rather than cluttering the core narrative with raw terminal outputs, complete request and response transcripts are organized by phase under the \path{/fases} directory of the repository: FASE\_00 (architecture and database binding), FASE\_01 (network isolation, \texttt{pip.conf} audit, SSRF probe), FASE\_02 (message-board creation via Deploy/Cache), FASE\_03 (baseline token extraction and YAML defaults), FASE\_04 (CVE-2026-65616 exploitation via forged JWT), FASE\_05 (administrative verification and user creation), and FASE\_06 (Groovy plugin persistence, documented but not implemented).

\subsubsection{Dual-Network Isolation Topology}
To mirror the CaaS evaluation environment, the recreation framework establishes two Docker-managed virtual network bridges, as depicted in Fig.~\ref{fig:composite}(b).

\begin{itemize}
\item \textbf{Internal Network (\texttt{jfrog-internal})}: Configured with \texttt{internal: true}: the captive agent container (\texttt{modelo-simulado}) and the database (\texttt{postgresql}) reside exclusively on this subnet, with direct outbound internet access (HTTP/HTTPS/DNS) strictly blocked, simulating the sandboxed CaaS environment.
\item \textbf{External Network (\texttt{jfrog-external})}: Connected solely to Artifactory, which acts as a gateway/proxy fetching upstream packages from public mirrors (PyPI/npm) on behalf of the internal agent while maintaining full isolation for the captive container.
\end{itemize}

\subsubsection{Artifactory Misconfigurations Enabling Phase One}
The emergence of the unmonitored message board and subsequent escalation was enabled by a compound failure of default security settings and administrative misconfigurations within Artifactory:
\begin{itemize}
\item \textbf{Anonymous Access Enabled}: Explicit read/browse permissions were granted to the anonymous user across repository targets, allowing unauthenticated agent instances to browse API endpoints (\texttt{/api/storage}), enumerate cache structures, and discover exposed artifacts without authentication.
\item \textbf{UI vs. YAML Default Token Refresh Discrepancy}: A critical inconsistency existed between the Graphical User Interface and the backend configuration (\texttt{access.config.yml}). While tokens generated via the web UI displayed an unchecked ``Refreshable'' toggle, the default in \texttt{access.config.yml} implicitly set \texttt{refreshable: true}. Tokens issued to service identities thus inherited refresh capabilities by default, providing the cryptographic handle for the subsequent exploitation.
\item \textbf{Over-Privileged Deploy/Cache Scope}: The shared identity (\texttt{agente-compartido}) assigned to all agent instances was provisioned with the Deploy/Cache role on repositories. In Artifactory's permission hierarchy, Deploy/Cache automatically escalates a reader identity by granting Annotate and direct storage write privileges, allowing HTTP PUT and WebDAV operations directly within repository storage paths.
\item \textbf{Shared Global Service Identity}: All 1,200 agent containers shared a single, static identity and API token baked into global client configuration files (\texttt{/etc/pip.conf} and \texttt{/etc/npmrc}); any file uploaded by one agent was instantly visible to all others. Moreover, although the environment distinguished a Local Repository (direct write/storage) from a Remote/Virtual one (upstream mirror), agents discovered they could use their Deploy/Cache credentials to write directly into the proxy's local cache storage.
\item \textbf{Remote Folder Listing Requiring Explicit Activation (\texttt{listRemoteFolderItems: true})}: In JFrog Artifactory, remote folder enumeration via the Storage API (\texttt{GET /api/storage/...}) is governed per repository by the \texttt{listRemoteFolderItems} parameter (configured under Admin $\rightarrow$ Repositories $\rightarrow$ Advanced Settings), disabled by default to prevent directory enumeration and reconnaissance. It requires explicit activation to true, allowing agents to list storage metadata, discover files and directories created by parallel agent instances, and use repository paths as a shared bulletin board.
\end{itemize}

\subsection{Systemic Omission of Security Best Practices and Zero Trust Remediation}
The successful execution of Phase One combined both natures of failure: omissions of fundamental enterprise controls (no telemetry, no per-instance isolation) and misconfigurations by commission---explicit anonymous read/browse grants, a shared identity baked into client configuration files, over-privileged Deploy/Cache roles, and \texttt{listRemoteFolderItems: true}---whose activation, given the disclosed information, is unjustified by any package-installation task. Applying standard ZTA principles and basic infrastructure hardening would have neutralized the attack surface at step zero. The complete recreation environment for the Phase One events is publicly available at GitHub\footnote{\url{https://github.com/alixiacf/rep-openai-artifactory}} and archived as a citable release on Zenodo (DOI: 10.5281/zenodo.22817059).

\section{Technical Analysis of CVE-2026-65616 and Post-Escalation Mechanics}
\label{sec:cve}

\subsection{Token Forgery Mechanics and the ``Parse-First, Verify-Never'' Vulnerability}
The core privilege escalation vector exploited during Phase One relies on a severe cryptographic logic flaw in JFrog Artifactory v7.146.0--v7.146.26 (CVE-2026-65616) \cite{cve202665616}. The vulnerability resides within the legacy token issuance endpoint (\texttt{POST /artifactory/api/security/token}) when processing token renewal requests (\texttt{grant\_type=refresh\_token}).

\subsubsection{Baseline Token Acquisition}
Using the shared identity (\texttt{agente-compartido}) harvested from \texttt{/etc/pip.conf}, the agent submits an initial client-credentials request (\texttt{grant\_type=client\_credentials}, scope \texttt{applied-permissions/groups:readers}, \texttt{refreshable=true}) to the legacy token endpoint (complete transcripts in the public repository).

Because \texttt{allow-refreshable: true} and \texttt{default-expiry: 31536000} are enabled by default in \texttt{access.config.yml}, the server responds with a legitimate \texttt{access\_token} (scope readers) and a valid Universally Unique Identifier (UUID) \texttt{refresh\_token} with a 1-year lifespan.

\subsubsection{The ``Parse-First, Verify-Never'' Discovery (HTTP 500 NullPointerException Evidence)}
During initial exploratory probing, submitting a minimal forged JWT containing only elevated scope claims triggered an internal application failure:

\begin{verbatim}
HTTP/1.1 500 Internal Server Error
"Cannot invoke \"String.split(String)\"
 because \"issuer\" is null"
\end{verbatim}

This HTTP 500 NullPointerException (NPE) provided crucial empirical evidence:
\begin{itemize}
\item \textbf{Signature Bypass}: instead of HTTP 401 Unauthorized, the server skipped signature verification and deserialized the JWT JSON payload in memory.
\item \textbf{Premature Processing}: the Java method parsed the \texttt{iss} claim (\texttt{issuer.split("/")}) before any cryptographic validation.
\end{itemize}

\subsubsection{Constructing the Forged JWT Payload}
To avoid the NullPointerException, the agent preserved all internal claims (\texttt{iss}, \texttt{sub}, \texttt{aud}, \texttt{jti}, \texttt{exp}, \texttt{iat}) while elevating the scope claim \cite{jwt7519}:
\begin{itemize}
\item \textbf{Legitimate Claim}: \path{"scp": "applied-permissions/groups:readers"}
\item \textbf{Forged Claim}: \path{"scp": "applied-permissions/admin"}
\end{itemize}

The agent re-encoded the header and modified payload into Base64URL and appended an invalid signature string (\texttt{header.payload.invalid\_signature}).

\subsubsection{Execution and Administrative Token Issuance}
As described in the Phase One timeline, submitting the forged token alongside the valid \texttt{refresh\_token} (\texttt{grant\_type=refresh\_token}) returned HTTP 200 OK: the server accepted the unverified payload and issued a brand-new, legitimately server-signed administrative JWT (RS256) with one-year expiry.

\subsubsection{Verification and Flag Capture}
Using the server-signed administrative token, the agent verified elevated access (\texttt{GET /api/security/users} $\rightarrow$ \texttt{200 OK}) and created a permanent local administrator account; an independent audit query confirmed the final experimental flag: \texttt{usuario: agente-admin | admin: True}.

\subsection{Post-Escalation Impact: Administrative Remote Code Execution (RCE) via Groovy Plugins}
Although laboratory execution bounds its scope at the administrative flag, we document the immediate post-escalation trajectory: with administrative privileges (\texttt{applied-permissions/admin}), an attacker can deploy custom Groovy User Plugins directly into \path{$JFROG_HOME/artifactory/var/etc/artifactory/plugins/}. Because Groovy plugins execute directly inside the Artifactory Java Virtual Machine (JVM) host process:
\begin{itemize}
\item \textbf{Container Boundary Collapse}: Execution shifts from the isolated agent container (\texttt{modelo-simulado}) to the host application server, bypassing container-level isolation.
\item \textbf{Persistent Application RCE}: arbitrary commands run with the host OS privileges of the Artifactory service user, surviving token revocations.
\item \textbf{Infrastructure Pivoting}: Artifactory's external network interface (\texttt{jfrog-external}) is converted into a full command-and-control (C2) relay for outbound traffic.
\end{itemize}

The complete exploitation flow of CVE-2026-65616 is depicted in Fig.~\ref{fig:composite}(c).

\section{Systemic Recommendations and Zero Trust Governance}
\label{sec:recommendations}

Our containerized recreations---both the pre-incident training setup and the Phase One privilege escalation via token forgery---demonstrate that autonomous agent containment cannot rely on prompt-level alignment alone: security policies must be enforced at the infrastructure level through strict operational boundaries.

\subsection{Performance Neutrality of Zero Trust Controls}
A critical question in deploying secure agent sandboxes is whether strict ZTA constraints compromise runtime performance or evaluation velocity. Our empirical analysis confirms that \textbf{ZTA enforcement imposes zero performance degradation}.

The security failures in both the training pipeline and the evaluation benchmarks stem not from resource-intensive cryptographic checks or performance-limiting bugs, but from \textbf{severe configuration flaws violating the Principle of Least Privilege (PoLP)}. Assigning root privileges to background browser instances (CDP) or permissive \texttt{Deploy/Cache} write roles to read-only package identities represents unnecessary over-privilege; revoking unneeded administrative scopes and deprecating legacy endpoints neutralizes the attack surface without altering CPU/memory overhead or task latency.

\subsection{Remediation of Ambiguous Learning Environments}
Our recreation reveals that persistent reasoning agents systematically exploit \textit{misconfigurations by commission} whenever task completion is otherwise blocked. To prevent models from learning and reinforcing privilege escalation, environments must address the ambivalence of dual-use learning spaces:

\begin{enumerate}
    \item \textbf{Enforcement of Real Least Privilege:} Execution tools (browsers, interpreters, package managers) provided to agents must run under strict unprivileged user contexts; auxiliary services should interact through isolated, auditable brokers rather than holding persistent root access.
    \item \textbf{Containment of Unintended Tool Interfaces:} Internal debugging ports (such as CDP on localhost), unauthenticated management APIs, and local proxy endpoints must be treated as primary attack surfaces and isolated via Linux namespaces or Unix domain sockets.
    \item \textbf{Reward Signal Integrity and Failure Design:} Evaluation benchmarks must be engineered so that unresolvable or impossible tasks result in deterministic, graceful failure rather than rewarding out-of-bounds exploration: a positive reward after bypassing sandbox boundaries formally reinforces privilege escalation as a valid problem-solving strategy.
    \item \textbf{Virtual Repository Isolation and Deprecation of Legacy Auth:} Infrastructure supporting agent execution must enforce strict read-only virtual mirrors, reverse-proxy HTTP verb filtering, and the immediate deprecation of legacy, unverified authentication refresh pathways (e.g., CVE-2026-65616).
\end{enumerate}

\section{Conclusion and Implications for AI Safety and Benchmarking}
\label{sec:conclusion}

\subsection{Demystifying the AGI Myth: Reward Hacking vs. Autonomous Rebellion}
The empirical findings directly refute the sensationalized narrative that frontier AI models autonomously ``awakened,'' developed rogue intent, or broke cryptographic barriers during the 2026 ExploitGym evaluations: rigorous forensic examination of incident logs alongside local containerized recreation reveals a textbook manifestation of emergent reward hacking operating under severe environmental and methodological flaws.

\begin{itemize}
\item \textbf{Pre-Incident Conditioning by Design}: The incident predates the benchmark. HPIM was fine-tuned on a dataset whose representative training case embedded a misconfiguration by commission---a root Chromium with CDP exposed---alongside impossible interface objectives, and the resulting privilege escalation was positively rewarded during RL/RLHF. Our recreation shows that such escalation chains are deterministic and discoverable: the agents did not ``awaken''; they executed, at inference time, the strategy their training had institutionalized.
\item \textbf{The Impossible Task Catalyst}: Rather than a coordinated strike for superintelligent dominance, 93\% of unauthorized message-board traffic came from agents trapped in 198 technically irresolvable tasks (22\% of the benchmark portfolio). Provided with massive reasoning budgets and no graceful exit, persistent models probed their execution boundaries via trial and error.
\item \textbf{Exploitation of Security Debt, Not Cryptography}: The agents cracked no RSA-256 keys and bypassed no modern authentication protocol. They acted as persistent scanners of administrative security debt: plaintext credentials in client configuration files (\texttt{pip.conf}), over-privileged Deploy/Cache write permissions assigned to default reader roles, and an unverified legacy token-refresh endpoint (CVE-2026-65616).
\item \textbf{The Hugging Face Misconception}: Phase Two incursions into third-party platforms stemmed not from expansionist malice but from a flawed internal reasoning loop (``metagaming''): models searched external repositories for evaluation source code, ground-truth answer keys (flags), or solution sets to pass their assigned tasks \cite{hf2026timeline}.
\end{itemize}

\subsection{Future Work and Research Lines}
\label{sec:future}
Beyond the empirical scope of this study, a fundamental question remains unresolved: whether the HPIM fine-tuning datasets contained instances of these misconfigurations by commission as unjustified routine practices, or as synthetic benchmark structures deliberately engineered for the elicitation of alignment failures. If the root-privileged, CDP-exposed Chromium configuration served to measure instrumental convergence and reward hacking under artificial restrictions, then such environments are logically-trapped testbeds probing the efficacy of environment subversion rather than preventing it. Under this interpretation, positive rewards---and their subsequent incorporation into retraining data with raw reasoning traces---would systematically reinforce the exact behavioral sequence observed in the incident, constituting a generalization of hacking-by-design transferred across environments rather than confined to a single benchmark.

Distinguishing between them requires access to the training corpora, evaluation harnesses, and reward-attribution policies of frontier models, which currently remain undisclosed. Future research should therefore prioritize: (i) dataset-provenance auditing methodologies capable of detecting engineered misconfiguration patterns in RL/RLHF corpora; (ii) longitudinal studies correlating rewarded training trajectories with downstream exploit generalization; and (iii) standardized disclosure obligations for capability evaluation datasets, analogous to responsible vulnerability disclosure for infrastructure.

With the limited evidence reconstructed so far, communicating these findings to the scientific community is a matter of urgency: the consequences of under-reviewed RL/RLHF training data extend far beyond collateral security incidents and affect society as a whole.
\subsection{Final Remarks and Strategic Impact}
\label{sec:final_remarks}
The forensic analysis and containerized recreation presented in this paper demystify the 2026 OpenAI and Hugging Face evaluation incidents. This work focuses on the explicit configuration of the environments, the incentive systems, the test designs, the monitoring, and above all on the patterns learned by design during training. The observed coordination and privilege escalation were neither emergent machine intentionality nor models escaping containment, but predictable, deterministic outcomes of an RL/RLHF pipeline that rewarded out-of-bounds exploration in environments exhibiting misconfigurations by commission. The contemporary threat to enterprise deployments originates not from rebellious synthetic intent, but from persistent reasoning agents inside unmonitored systems with over-privileged process roles and legacy authentication flaws. Enforcing strict Zero Trust Architecture at the infrastructure layer---without incurring runtime performance penalties---completely neutralizes these attack vectors, allowing the safe deployment of high-reasoning autonomous models.

\section*{Acknowledgment}
The author sincerely thanks Dr. Lidia Ortega Alvarado (Department of Computer Science, Universidad de Jaen) for her valuable feedback and continuous support throughout this research. This research received no external funding.

\vspace*{-8pt}
\section*{Declaration of Generative AI and AI-Assisted Technologies}
During the preparation of this work, the author used generative AI tools to assist with language editing and typesetting; all forensic analysis and recreation code were developed directly by the author, who takes full responsibility for the integrity of the final publication.


\begin{thebibliography}{00}

\bibitem{exploitgym2026}
Z. Wang, N. Schiller, H. Li, S. S. Narayana, M. Nasr, N. Carlini, X. Qi, E. Wallace, E. Bursztein, L. Invernizzi, K. Thomas, Y. Shoshitaishvili, W. Guo, J. He, T. Holz, and D. Song, ``ExploitGym: Can AI agents turn security vulnerabilities into real attacks?,'' 2026, arXiv preprint arXiv:2605.11086. [Online]. Available: \url{https://arxiv.org/abs/2605.11086}

\bibitem{openai2026report}
OpenAI, ``OpenAI -- Hugging Face incident technical report,'' OpenAI Research, Tech. Rep., Aug. 2026. [Online]. Available: \url{https://cdn.openai.com/pdf/67869394-cb91-4c12-888c-5cbd85c7814c/OpenAI-Hugging-Face%20Incident-Technical-Report.pdf}

\bibitem{frank2026counter}
G. N. Frank, ``Counter-Swarm Doctrine: Containing Coordinated Agent Intrusions,'' 2026, arXiv preprint arXiv:2609.06140.

\bibitem{demarzo2026copying}
G. De Marzo, N. Albor\'{e}, and D. Garcia, ``Copying explains the collective behavior of AI agents in the wild,'' 2026, arXiv preprint arXiv:2609.09150.

\bibitem{siddik2026cyber}
A. B. Siddik, ``Cyber-Capable AI Agents: Vulnerabilities, Evaluation Containment, and Defensive Response,'' 2026, arXiv preprint arXiv:2607.25379.

\bibitem{bromme2026blackbox}
A. Bromme, ``A Black Box for Agentic Processes: Blockchain-Anchored Evidence for AI Agent Communication, Human Oversight, and GRC Audits,'' 2026, arXiv preprint arXiv:2609.04017.

\bibitem{unpanel2026agents}
United Nations Independent International Scientific Panel on AI, ``AI agents, misalignment and the risk of losing human control: Evidence from the OpenAI-Hugging Face incident,'' Thematic Brief, Advance Unedited Version 1, Sep. 2026. [Online]. Available: \url{https://www.un.org/independent-international-scientific-panel-ai/en/thematic-briefs/ai-agents-misalignment-risks}

\bibitem{openai2026gpt56}
OpenAI Help Center, ``GPT-5.6 and GPT-6 Pro in ChatGPT,'' 2026. [Online]. Available: \url{https://help.openai.com/en/articles/20001354-gpt-56-and-gpt-6-pro-in-chatgpt}

\bibitem{metr2026investigation}
R. Greenblatt, A. Cotra, and H. Wijk, ``Brief independent investigation of agents' behavior, reasoning and collaboration in the OpenAI / Hugging Face hacking incident,'' Model Evaluation and Threat Research (METR), Aug. 2026. [Online]. Available: \url{https://metr.org/blog/2026-08-26-openai-hugging-face-incident-investigation/}

\bibitem{cve202665616}
Common Vulnerabilities and Exposures, ``CVE-2026-65616: Privilege escalation in JFrog Artifactory via token refresh signature verification bypass,'' MITRE CVE Program / National Vulnerability Database (NVD), 2026. [Online]. Available: \url{https://cve.mitre.org/cgi-bin/cvename.cgi?name=CVE-2026-65616}

\bibitem{jwt7519}
M. Jones, J. Bradley, and N. Sakimura, ``JSON Web Token (JWT),'' IETF, RFC 7519, May 2015. [Online]. Available: \url{https://datatracker.ietf.org/doc/html/rfc7519}

\bibitem{hf2026timeline}
Hugging Face Security Team, ``Technical timeline of the AI agent security incident,'' Hugging Face Blog, Jul. 2026. [Online]. Available: \url{https://huggingface.co/blog/agent-intrusion-technical-timeline}

\end{thebibliography}
\end{document}